\documentclass[english,acus]{JAC2003}
\usepackage[T1]{fontenc}
\usepackage[latin9]{inputenc}
\usepackage{amsmath}
\usepackage{graphicx}

\makeatletter

\newcommand{\noun}[1]{\textsc{#1}}
\providecommand{\tabularnewline}{\\}

\usepackage{array}

\makeatother

\usepackage{babel}
\begin{document}

\title{CRAB CAVITY IN CERN SPS}

\author{H. J. Kim\thanks{hjkim@fnal.gov} and T. Sen, FNAL, Batavia, IL 60510,
USA}
\maketitle
\begin{abstract}
Beam collisions with a crossing angle at the interaction point have
been applied in high intensity colliders to reduce the effects of
parasitic collisions which induce emittance growth and beam lifetime
deterioration. The crossing angle causes the geometrical reduction
of the luminosity. Crab cavity can be one of the most promising ways
to compensate the crossing angle and to realize effective head-on
collisions. Moreover, the crab crossing mitigates the synchro-betatron
resonances due to the crossing angle. Crab cavity experiment in SPS
is proposed for deciding on a full crab-cavity implementation in LHC.
In this paper, we investigate the effects of crab crossing on beam
dynamics and its life time with the global scheme.
\end{abstract}

\section{INTRODUCTION}

In order to achieve a high luminosity in high-field colliders with
standard collision schemes, one can consider small bunch spacings
inside the beam or a beam current increase. However, the increase
of beam intensity may result in beam instability and high beam power
loss. A number of bunches inside the beam bring into parasitic encounters
at the interaction region which cause emittance growth and beam loss.
Increasing the crossing angle to reduce the effects of parasitic collisions
causes luminosity reduction by exciting the synchro-betatron resonance. 

A crab crossing scheme has been proposed to allow a large crossing
angle for both linear colliders and circular colliders without a loss
of luminosity \cite{Palmer}. When a particle is passing through a
crab cavity structure, it not only increases the longitudinal energy,
but also changes its transverse momentum. The crab cavity can compensate
the horizontal or vertical crossing angle at the interaction point.
The two counter-moving beams, therefore, experience an effective head-on
collision. For the LHC upgrade phase II, both local and global crab
crossing schemes are under consideration. At present a possible configuration
is to install only one global crab cavity in the IR4 section because
a minimum separation of beam lines is required by the RF structure
\cite{Brunner}. In addition, a compact cavity suitable for the two
beam separation is under development.

We consider the experiment in the CERN SPS with KEK-B crab cavity,
which has several advantages. Since the beam parameter of SPS is close
to that of the LHC, the SPS is the best test bed for the LHC crab
crossing. The KEK-B crab cavity could be available after KEK-B physics
run, and installed in the SPS Coldex location. In this paper, we investigate
the effects of crab crossing on beam dynamics and its life time with
the global scheme in the SPS with KEK-B crab cavity using six dimensional
weak-strong tracking code \noun{bbsimc} \cite{Kim}.

\section{MODEL}

\begin{figure}
\begin{centering}
\includegraphics[bb=25bp 15bp 340bp 270bp,clip,scale=0.725]{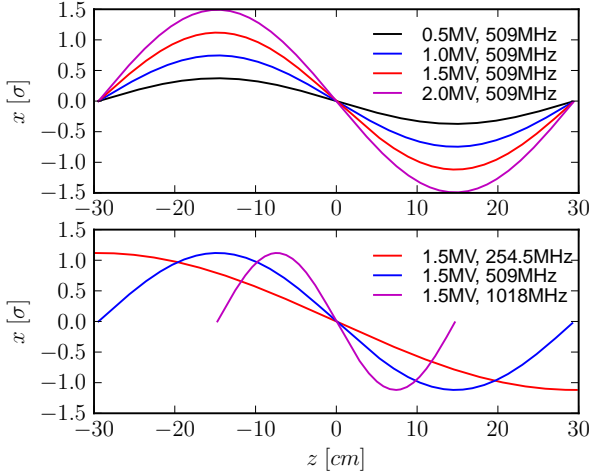}
\par\end{centering}

\caption{Closed orbit at the beginning of the SPS lattice due to a crab-cavity
kick: (top) crab cavity voltage, and (bottom) crab cavity frequency.
The closed orbit is a function of longitudinal position inside a bunch.
The full wave length of crab cavity is $\lambda_{cc}$ = 58.9 cm.
\label{fig:Closed-orbit}}
\end{figure}

\begin{table}
\caption{SPS optics parameters at two locations of crab cavity and cavity parameters.
\label{tab:SPS-table}}

\centering{}%
\begin{tabular}{c|c|c}
\hline 
Quantity & Coldex & Zero\tabularnewline
\hline 
location & 4010 m & 4094 m\tabularnewline
length & \multicolumn{2}{c}{10.272 m}\tabularnewline
$\left(\beta_{x},\beta_{y}\right)$ & (30.3, 76.8) & (94.0, 23.5)\tabularnewline
$\left(\mu_{x},\mu_{y}\right)$ & (15.173, 15.176) & (15.477, 15.497)\tabularnewline
$\left(D_{x},D_{y}\right)$ & (-0.476, 0.0) & (0.0, 0.0)\tabularnewline
voltage & \multicolumn{2}{c}{1.5 MV}\tabularnewline
frequency & \multicolumn{2}{c}{509 MHz}\tabularnewline
\hline 
\end{tabular}
\end{table}
 In case of a horizontal crossing, the kicks from the crab cavity
are given by \cite{Sun}\begin{subequations}\label{eq:MD-1} 
\begin{align}
\Delta x^{\prime} & =-\frac{qV}{E_{0}}\sin\left(\phi_{s}+\frac{\omega z}{c}\right),\label{eq:MD-1a}\\
\Delta\delta & =-\frac{qV}{E_{0}}\cos\left(\phi_{s}+\frac{\omega z}{c}\right)\cdot\frac{\omega}{c}x,\label{eq:MD-1b}
\end{align}
 \end{subequations} where $q$ denotes the particle charge, $V$
the voltage of crab cavity, $E_{0}$ the particle energy, $\phi_{s}$
the synchronous phase of the crab-cavity rf wave, $\omega$ the angular
rf frequency of the crab cavity, $c$ the speed of light, $z$ the
longitudinal coordinate of the particle with respect to the bunch
center, and $x$ the horizontal coordinate. 
\begin{figure}
\begin{centering}
\includegraphics[bb=20bp 15bp 340bp 270bp,clip,scale=0.725]{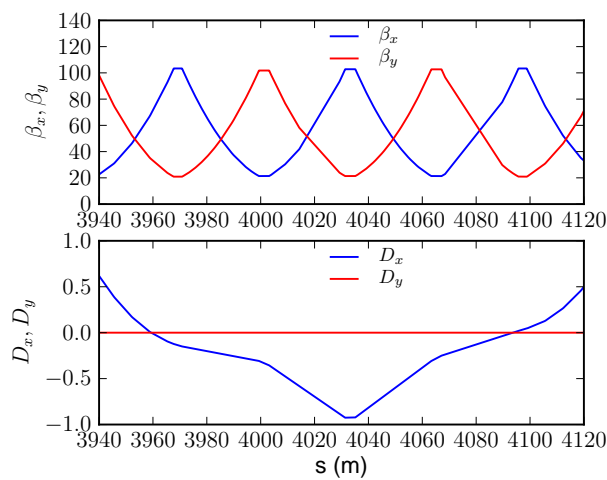}
\par\end{centering}

\caption{Twiss function in crab cavity location. \label{fig:Twiss}}
\end{figure}
 The global crab cavity causes a closed orbit distortion dependent
on the longitudinal position of particles, and has the equilibrium
beam envelope tiled all around the ring. For a small bunch less than
the rf wavelength of the crab cavity deflecting mode, the tilt angle
of the beam envelope is given by 
\begin{align}
\tan\theta_{crab} & =\frac{qV\omega\sqrt{\beta\beta_{crab}}}{c^{2}p_{0}}\left|\frac{\cos\left(\Delta\varphi-\pi Q\right)}{2\sin\pi Q}\right|,\label{eq:MD-2}
\end{align}
 where $\beta$ is the beta function at the BPM position, $\beta_{crab}$
the beta function at the crab cavity, $\Delta\varphi$ the phase advance
between the crab cavity location and the BPM, and $Q$ the betatron
tune. The angle $\theta_{crab}$ is proportional to the crab cavity
voltage and frequency, and the beta functions. Figure \ref{fig:Closed-orbit}
shows the closed orbit at the beginning of the SPS optics which is
a sinusoidal function versus the full wave length of crab cavity.
 The slope at $z=0$ is related to the tilt angle of the beam envelope.

\section{RESULTS}

\begin{figure}
\begin{centering}
\includegraphics[scale=0.65]{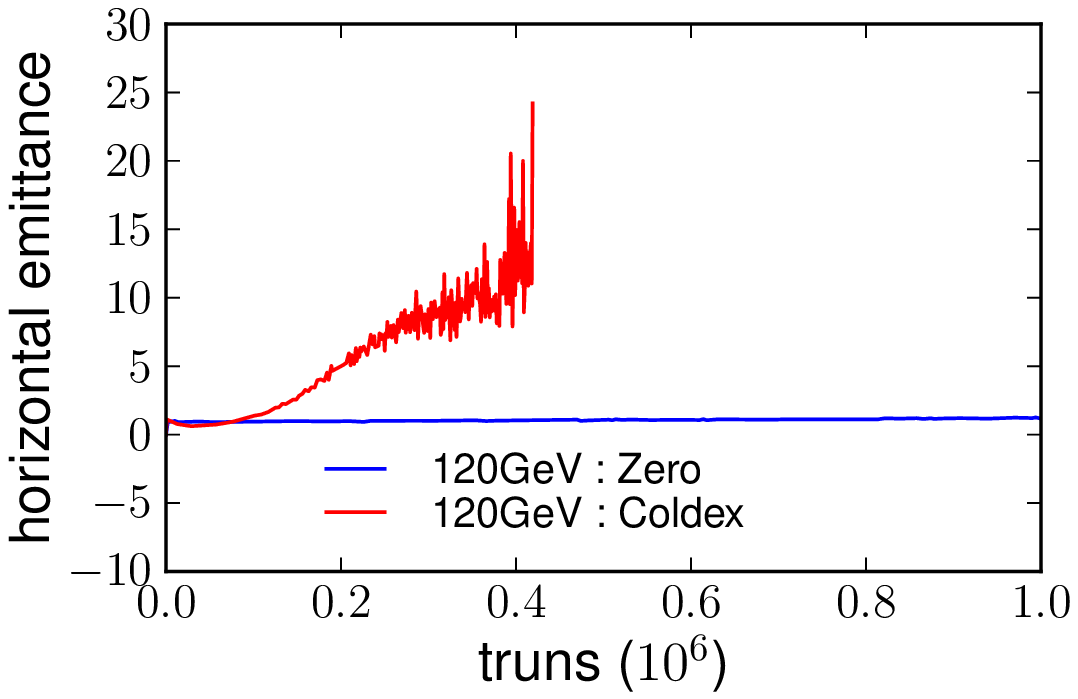}
\par\end{centering}

\begin{centering}
\includegraphics[scale=0.65]{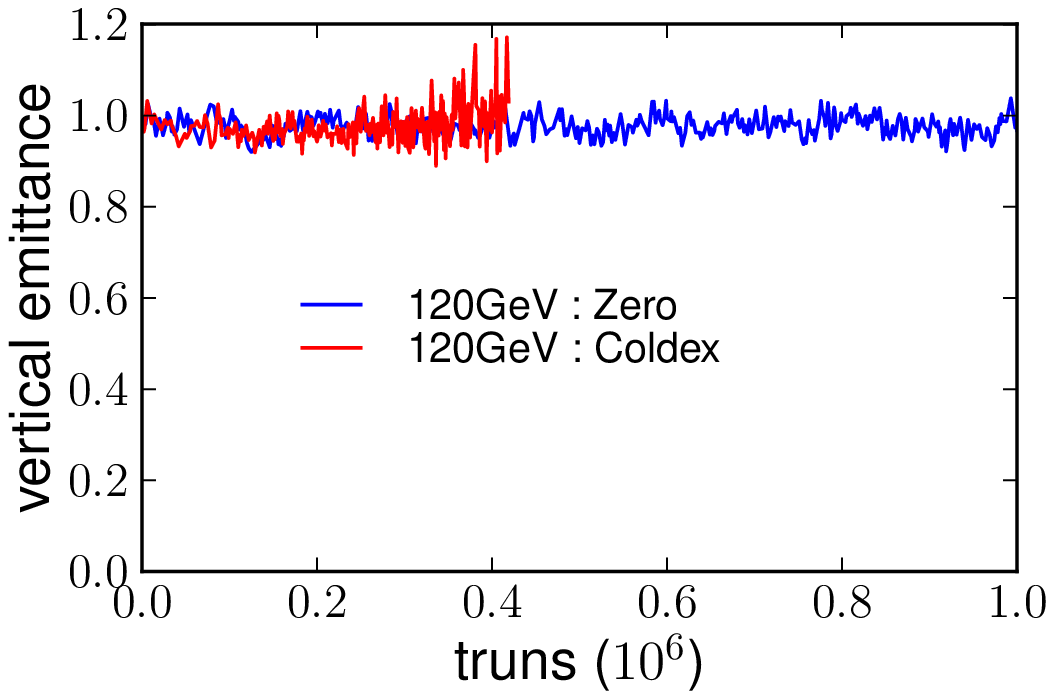}
\par\end{centering}

\begin{centering}
\includegraphics[scale=0.65]{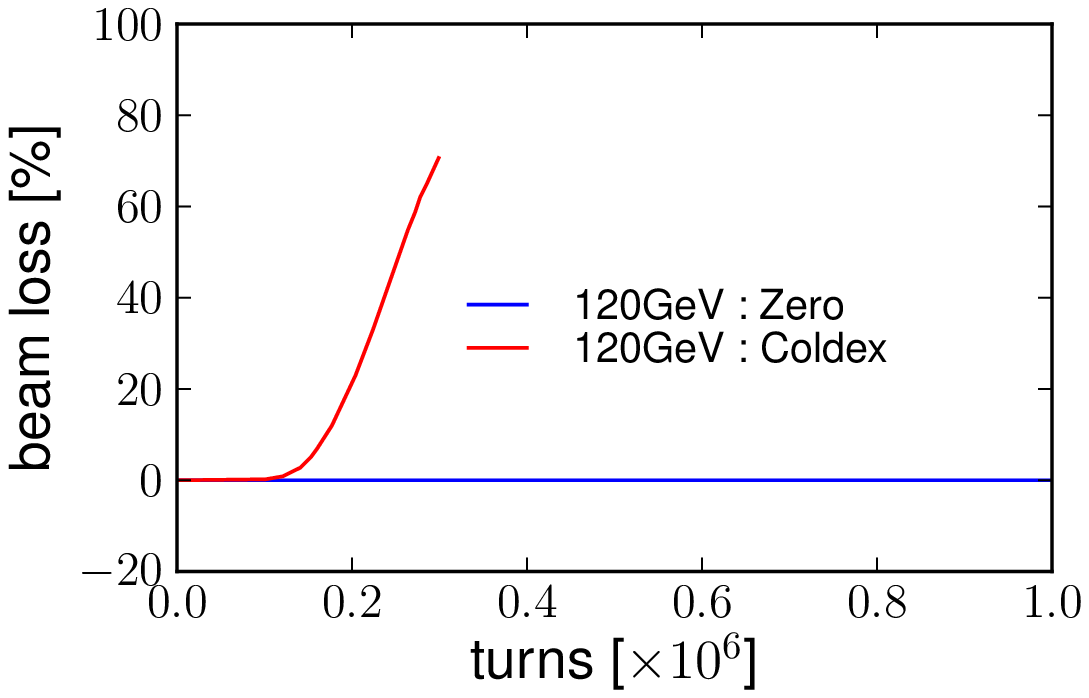}
\par\end{centering}

\caption{Transverse emittance and beam loss for 120GeV SPS beam when a crab
cavity is installed at two different locations. \label{fig:two-locs}}
\end{figure}

\begin{figure}
\begin{centering}
\includegraphics[bb=0bp 0bp 361bp 216bp,scale=0.65]{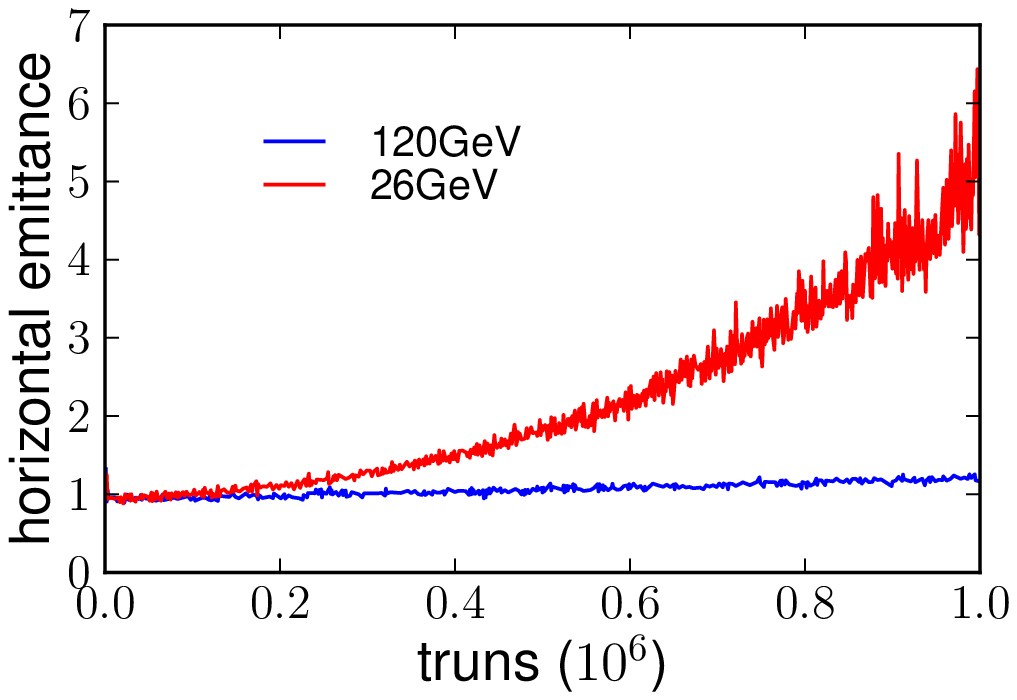}
\par\end{centering}

\begin{centering}
\includegraphics[scale=0.675]{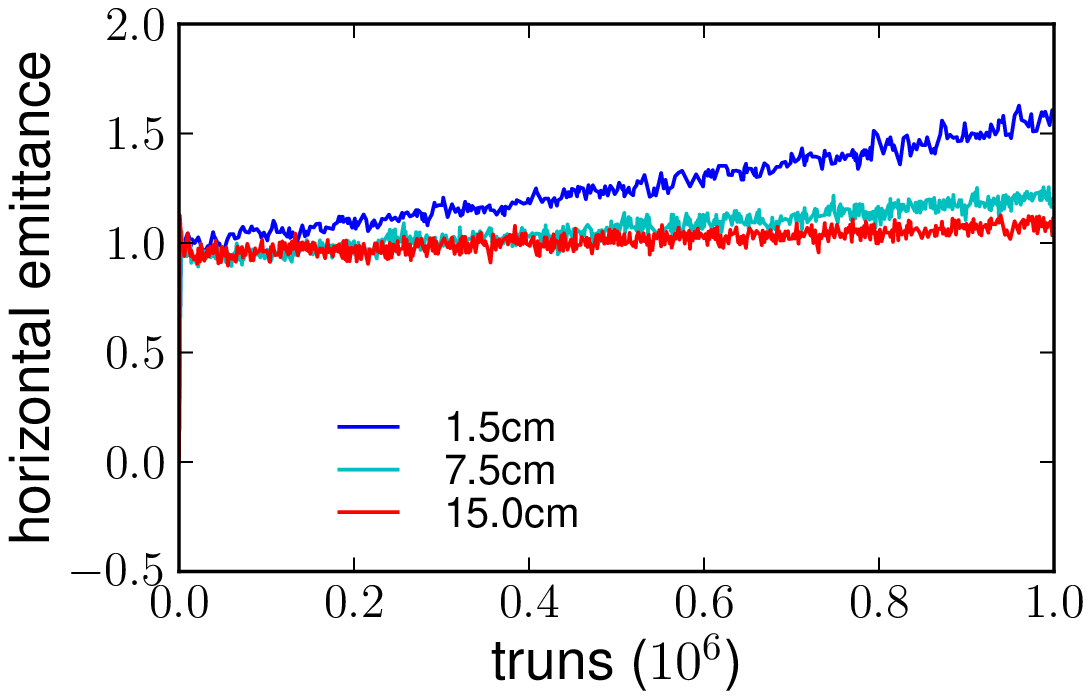}
\par\end{centering}

\caption{Transverse emittance variations for different energy (top) and different
bunch length (bottom). Crab cavity is installed at Zero location.
\label{fig:Zero}}
\end{figure}

\begin{figure}
\begin{centering}
\includegraphics[scale=0.65]{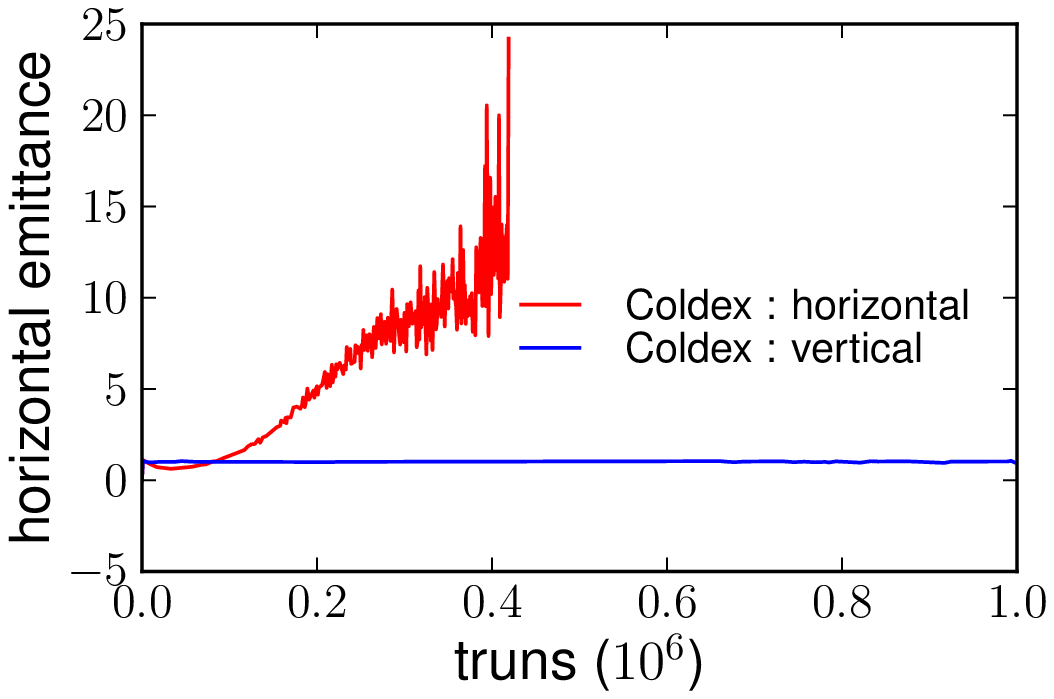}
\par\end{centering}

\begin{centering}
\includegraphics[scale=0.65]{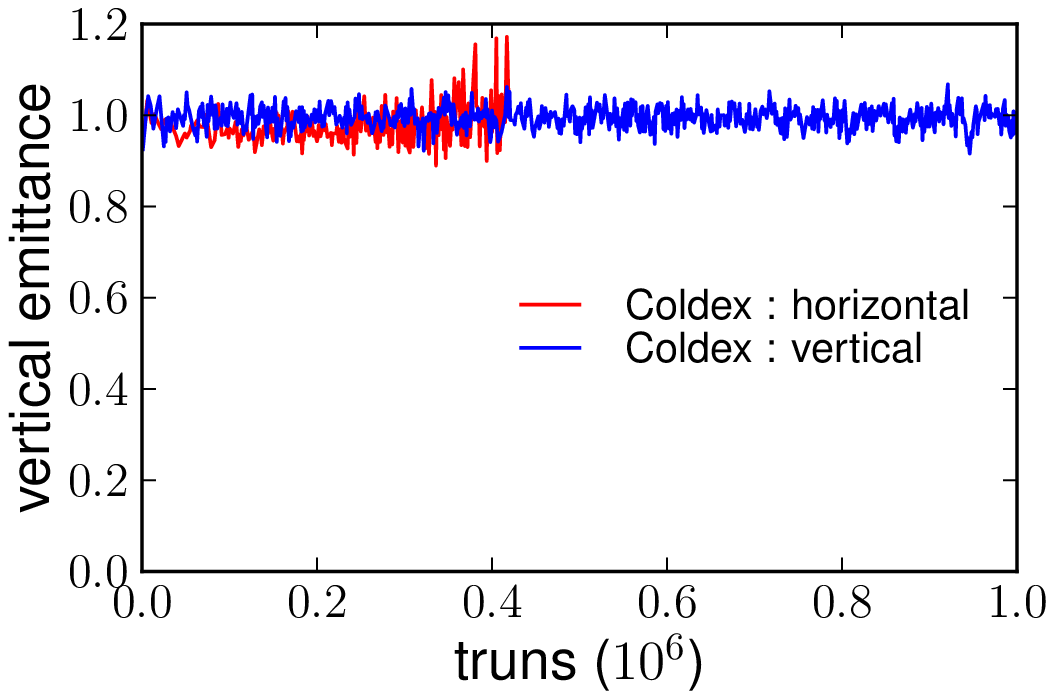}
\par\end{centering}

\caption{Transverse emittance and beam loss of 120GeV SPS beam for horizontal
and vertical crossings. Crab cavity is installed in Coldex location.
\label{fig:vertical}}
\end{figure}

The SPS optics parameters are listed in Table \ref{tab:SPS-table}.
The Coldex Cryogenic Experiment location is proposed for an installation
of KEK-B crab cavity of which maximum voltage and frequency is 1.5
MV and 509 MHz respectively. The full wave length of crab cavity is
$\lambda_{cc}$ = 58.9 cm. Figure \ref{fig:two-locs} shows the transverse
emittance and beam loss for 120 GeV SPS beam when a crab cavity is
installed at Coldex location. 
 It is observed from the simulation results that the crab cavity leads
to blow-up of the horizontal beam emittance and significant particle
losses while the vertical emittance keeps constant. In order to see
the effects of crab cavity on beam dynamics, consider a synchronous
particle, i.e., $z=\delta=0$. The horizontal and longitudinal momentum
due to the crab cavity is $\Delta x^{\prime}=0$, and $\Delta\delta=-\left(qV\omega/E_{0}c\right)x$.
After a first turn, one can get 
\begin{align}
\left(\begin{array}{c}
x\\
x^{\prime}
\end{array}\right)_{1} & =\left(\mathbf{M}+\mathbf{D}\right)\left(\begin{array}{c}
x\\
x^{\prime}
\end{array}\right)_{0},\label{eq:SR-1}
\end{align}
 where $\mathbf{M}$ is the transfer matrix for a full revolution.
The periodic dispersion $\vec{\eta}=\left(\eta_{x},\eta_{x}^{\prime}\right)$
at the cavity and the cavity kick determine $\mathbf{D}=\left(\begin{array}{cc}
-\left(qV\omega/E_{0}c\right)\left(\mathbf{I}-\mathbf{M}\right)\vec{\eta} & \vec{0}\end{array}\right)$. The stability boundary can be obtained from the determinant of $\mathbf{M}+\mathbf{D}$,
\begin{align}
\eta_{x}(1-\cos\mu+\alpha_{x}\sin\mu)+\eta_{x}^{\prime}\beta_{x}\sin\mu & =0.\label{eq:SR-2}
\end{align}
 It is interesting to note that the condition depends on the twiss
functions of lattice, the crab cavity parameters, and the dispersion
functions at the crab cavity. Provided that the dispersion function
in the cavity is zero, there is no stop bend of the beam instability.
In order to see the effects of the dispersion, the new cavity location
where $\eta_{x}$ is zero is chosen. As shown in Figure \ref{fig:two-locs},
the horizontal emittance for a cavity at Zero location, i.e., $\eta_{x}=0$,
is much less than that for Coldex cavity. However, a finite increase
of horizontal emittance is observed because $\eta_{x}^{\prime}$ is
finite. Figure \ref{fig:Zero} shows the horizontal emittance for
different beam energy and bunch length. 
 The emittance increase is large for small beam energy, which is related
to a finite dispersion. For a fixed bunch length $\sigma_{z}=7.5$
cm, the energy spreads are $\Delta E$ = 37.2 MeV and $\Delta E$
= 51.6 MeV for 120 GeV and 26 GeV respectively. The coefficient of
momentum change due to the crab cavity becomes large when the beam
energy is small. Furthermore, if we look at the effects of bunch length
on the horizontal emittance, it is surprisingly found that the beam
with small bunch length is less stable. The transverse kick $\Delta x^{\prime}$
is approximately linear only over a small portion of the wave length
of crab cavity, for example, $\left|z\right|<\lambda_{cc}/6$. The
crab cavity distorts the beam envelope for a large longitudinal bunch.
The average change of longitudinal momentum due to crab cavity becomes
large for a short bunch because $\Delta\delta$ is proportional to
$\cos\frac{\omega}{c}z$. The finite dispersion effect, therefore,
seems stronger than that of nonlinearity due to the beam distortion.
In addition, it is a reasonable motivation to check how the vertical
crab cavity at the Coldex location gives effects on the vertical emittance
because the vertical dispersion function is zero for both $\eta_{y}$
and $\eta_{y}^{\prime}$ all over the ring at least in the lattice
optics. Obviously, we do not see any emittance growth in both horizontal
and vertical planes, as shown in Figure \ref{fig:vertical}. 

\section{SUMMARY}

In this paper, we investigate the effects of the global crab cavity
on the dynamic aperture and the transverse emittance growth. The results
show that the dispersion function at the crab cavity location matters.
Even a finite dispersion can induce the emittance growth. Dispersion
matching (both $\eta_{x}$and $\eta_{x}^{\prime}$) at the location
where the crab cavity is installed is required to minimize the beam
life time reduction. Vertical crab crossing scheme may be need to
be considered due to its small vertical dispersion.

\section*{ACKNOWLEDGMENTS}

This work was supported by the US Department of Energy through the
US LHC Accelerator Research Program (LARP).

\end{document}